
\input harvmac


\noblackbox
{\parskip=0pt
\baselineskip=22pt      

\Title{\vbox{\hbox{CTP-TAMU-95/91}}}
{\vbox{\centerline{Orbits of a String around a Fivebrane}}}

\centerline{Ramzi R. Khuri\footnote{$^\dagger$}{%
e-mail address:rrk@phys.tamu.edu, rrk@tamphys.bitnet}\
and HoSeong ~La\footnote{$^*$}{%
e-mail address: hsla@phys.tamu.edu, hsla@tamphys.bitnet}}

\bigskip\centerline{Center for Theoretical Physics}
\centerline{Texas A\&M University}
\centerline{College Station, TX 77843-4242, USA}
\vskip 0.6in
The classical orbits of a test string in the transverse space of a singular
heterotic fivebrane source are classified.
The orbits are found to be either circular or open, but not conic because
the inverse square law is not satisfied at long range.
This result differs from predictions of General Relativity.
The conserved total angular momentum contains an intrinsic component from
the fivebrane source, analogous to the electron-monopole case.
}
\baselineskip=22pt

\bigskip
{\noindent
PACS numbers: 11.17.+y, 04.60.+n  }

\Date{12/91} 
 \noblackbox
\baselineskip=22pt
 \def\foot##1{\ref\dummy{##1}}
 \def\listrefs{\footatend\vfill\eject\immediate\closeout\rfile%
 \writestoppt%
 \baselineskip=22pt\centerline{{\bf References}}\bigskip{\frenchspacing%
 \parindent=20pt\escapechar=` \input refs.tmp\vfill\eject}%
 \nonfrenchspacing}             

\def\hf{{1\over 2}} 

\def\d{{\rm d}}
\def\e{{\rm e}}
\def\pa{\partial}

\def\eps{\epsilon}
\def\lbr{\left(}
\def\rbr{\right)}
\font\cmss=cmss10 \font\cmsss=cmss10 scaled 833
\def\IZ{\relax\ifmmode\mathchoice
{\hbox{\cmss Z\kern-.4em Z}}{\hbox{\cmss Z\kern-.4em Z}}
{\lower.9pt\hbox{\cmsss Z\kern-.4em Z}}
{\lower1.2pt\hbox{\cmsss Z\kern-.4em Z}}\else{\cmss Z\kern-.4em Z}\fi}

\def\cos{{\rm cos}}
\def\sin{{\rm sin}}

\vfill\eject
%

\lref\duf{M.J. Duff, Class. Quan. Grav. {\bf 5} (1988) 189\semi
M.J. Duff, in {\it Superworld II}, ed. by A. Zichichi (Plenum, New York,
1990).}

\lref\monto{C. Montonen and D. Olive, Phys. Lett.{\bf 72B} (1977) 117.}

\lref\hali{J.A. Harvey and J. Liu, Chicago preprint, EFI-91-27.}

\lref\ccrk{C.~G.~Callan and R.~R.~Khuri, Phys. Lett. {\bf B261} (1991) 363.}

\lref\dghrr{A.~Dabholkar, G.~Gibbons, J.~A.~Harvey and F.~Ruiz Ruiz,
Nucl. Phys. {\bf B340} (1990) 33.}

\lref\dfluone{M.~J.~Duff and J.~X.~Lu, Nucl. Phys. {\bf B354} (1991) 141.}

\lref\dflutwo{M.~J.~Duff and J.~X.~Lu, Nucl. Phys. {\bf B354} (1991) 129.}

\lref\dfluthree{M.~J.~Duff and J.~X.~Lu, Phys. Rev. Lett. {\bf 66}
(1991) 1402.}

\lref\dflufour{M.~J.~Duff and J.~X.~Lu, Nucl. Phys. {\bf B357} (1991)
534.}

\lref\dflufive{M.~J.~Duff and J.~X.~Lu, {\it A Duality Between Strings and
Fivebranes}, CTP-TAMU-28/91 (to appear in Class. Quant. Gravity).}

\lref\dflusix{M.~J.~Duff and J.~X.~Lu, {\it The Self-Dual Type II-B
Superthreebrane}, CTP-TAMU-29/91 (to appear in Phys. Lett. {\bf B}).}

\lref\chsone{C.~G.~Callan, J.~A.~Harvey and A.~Strominger, Nucl. Phys.
{\bf B359} (1991) 611.}

\lref\chstwo{C.~G.~Callan, J.~A.~Harvey and A.~Strominger, Nucl. Phys.
{\bf B367} (1991) 60.}

\lref\strom{A.~Strominger, Nucl. Phys. {\bf B343} (1990) 167.}

\lref\dfst{M.~J.~Duff and K.~S.~Stelle, Phys. Lett. {\bf B253} (1991)
113.}

\lref\host{G.~T.~Horowitz and A.~Strominger, Nucl. Phys. {\bf B360}
(1991) 197.}

\lref\hlp{J.~Hughes, J.~Liu and J.~Polchinski, Phys. Lett. {\bf B180}
(1986).}

\lref\town{P.~K.~Townsend, Phys. Lett. {\bf B202} (1988) 53.}

\lref\duff{M.~J.~Duff, Class. Quant. Grav. {\bf 5} (1988).}

\lref\khlatwo{R.~R.~Khuri and H.S.~La, {\it String Motion in Fivebrane
Geometry}, Texas A\&M preprint, CTP-TAMU 98/91 (1991).}

\lref\dkl{M.~J.~Duff, R.~R.~Khuri and J.~X.~Lu {\it String and Fivebrane
Solitons: Singular or Non-singular?}, Texas A\&M preprint,
CTP-TAMU 89/91 (1991).}


One of the main difficulties for string theory to be a realistic theory of
nature is its lack of nonperturbative structures.
This obstacle, however, is probably not insurmountable.
An important lesson we
learned from gauge field theory is that nonperturbative structures of a
theory can be often provided by certain classical solitonic solutions, such as
instantons. With this motivation in mind, classical solitonic solutions of
various string theories have been actively investigated lately%
\ref\Fbrev{For recent reviews, see
M.J.~Duff and J.X. Lu, ``A Duality between Strings and Fivebranes,''
Texas A\&M preprint, CTP-TAMU-28/91 (1991);
C.G. Callan, J.A. Harvey and A. Strominger, ``Supersymmetric String Solitons,''
Chicago preprint, EFI-91-66 (1991); and references therein.}.

In recent work, Callan, Harvey and Strominger demonstrated that a certain
solution to the supergravity super-Yang-Mills equations
is in fact an exact fivebrane (five-dimensional extended object)
solution of heterotic
string theory\refs{\chsone,\chstwo}. The transverse space of a fivebrane
propagating in ten-dimensional space-time is a four-dimensional Euclidean
space, in which the fivebrane can be regarded as a point-like object.
This fivebrane solution has the structure of a soliton expressed in terms of
a Yang-Mills instanton in the transverse space.

In this letter we study the classical orbits
of a test string around this exact ``symmetric'' fivebrane with one of its
spatial directions parallel to the string, and try to make an analogy with
the electron-monopole system. In contrast to General Relativity
the dynamics of the string-fivebrane system is governed by the competition
between the attractive gravitational force and the repulsive force due to the
antisymmetric tensor.
A more detailed study which includes the case of the
gauge solution of ref.\strom\ will be presented elsewhere\refs\khlatwo.

For this purpose we
need only write down the explicit form of the massless fields:
the metric $g_{MN}$, dilaton $\phi$ and antisymmetric
tensor $B_{MN}$.\ref\ftone{We ignore the contribution from the worldsheet
fermions because they are coupled to
the instanton  Yang-Mills background $A_M$ so that we do not expect any
qualitative change in the dynamics.
On  transforming these fields to the $7$-form formulation
of supergravity, one recovers the fivebrane solution first found
in M.~J.~Duff and J.~X.~Lu, Nucl. Phys. {\bf B354} (1991) 141.
For this solution, the analysis is similar.}\
Since we are interested in studying classical
``stringy'' effects produced by the fivebrane,
it is most convenient to write down the solution for the massless fields
in terms of the string ``$\sigma$-model'' metric:
\eqn\fivebrane{\eqalign{e^{2\phi}&=1+{Q\over r^2},\cr
ds^2&=\eta_{\mu\nu}dx^\mu dx^\nu + e^{2\phi}\delta_{mn}dx^m dx^n,\cr
H_{mnp}&=\eps_{mnp}{}^q\pa_q\phi,\cr}}
where $H=dB$, $\mu,\nu=0,1,...,5$ are fivebrane indices, $m,n=6,7,8,9$
are transverse indices and $M,N=0,1,2,...,9$ will be used for
spacetime indices. $r$ is the radial coordinate in
the transverse space, and for convenience
we have set $\phi_0=0$ in  Eq.\fivebrane,
using the scale symmetry $\phi\to\phi+\ln c, \
r\to c^{-1}r$ for some constant $c$.
The quantization of the Wess-Zumino term in the $\sigma$-model action
implies $Q=n\alpha'$, where $n$ is an
integer. The above single fivebrane solution can be immediately generalized
to a multi-fivebrane solution by linear superposition of multiple sources.


We now wish to study the orbits of a test string in the background created
by a fivebrane of the form \fivebrane. The Lagrangian for a string moving
in a given background of massless fields is given by
\eqn\lag{{\cal L}=\hf\left(\sqrt{-\gamma}\gamma^{ij}\pa_i X^M\pa_j X^N g_{MN}
+ \eps^{ij}\pa_i X^M \pa_j X^N B_{MN}\right),}
where $\gamma_{ij}$ is the worldsheet metric for $(i,j)=(\tau,x)$ and
$g_{MN}$ is the string $\sigma$-model metric.  For simplicity we take
the string to be parallel to one of the fivebrane directions, i.e. we assume
$x$ will be identified with one of the fivebrane coordinates, but we expect
that the results will be qualitatively similar in other cases.
As a result, the test string also appears as a point-like object in the
transverse space.
Since $B_{MN}$ is only nonzero when both $M$ and $N$ are transverse,
it follows that the Wess-Zumino term in the above Lagrangian vanishes
for the above fivebrane background.

We substitute the worldsheet constraint equation
$\gamma_{ij}=\pa_i X^M \pa_j X^N g_{MN}$ in \lag\ so that the Lagrangian
reduces to
\eqn\elag{\CL={\sqrt{-\gamma}}=\left[{\dot t}^2-\e^{2\phi}\left\{{\dot r}^2
+r^2\left({\dot\chi}^2+\sin^2\chi({\dot \theta}^2+\sin^2\theta{\dot\varphi}^2)
\right)\right\}\right]^{1/2},}
where $(r,\chi,\theta,\varphi)$ are spherical coordinates in the four
dimensional transverse space and the time derivative ``$\cdot$'' is taken with
respect to
the proper time $\tau$. Using four-dimensional spherical symmetry, we take
$\chi=\theta=\pi/2$ so that the problem reduces to a two-dimensional one with
a simplified Lagrangian
\eqn\lagthree{{\cal L}=\lbr\dot t^2-e^{2\phi}\lbr\dot r^2+r^2\dot \varphi^2\rbr
\rbr^{1/2}.}

For time-like geodesics, ${\cal L}=1$. Along these geodesics,
the Lagrangian \lagthree\ has two constants of motion
\eqn\com{\eqalign{E&\equiv {\partial {\cal L}\over \partial\dot t}=
{\dot t\over {\cal L}}=\dot t ,\cr
L&\equiv -{\partial {\cal L}\over \partial\dot\varphi}={e^{2\phi}r^2\dot\varphi
\over {\cal L}}=e^{2\phi}r^2\dot\varphi , \cr}}
where $E\geq 1$ is the conserved energy per unit
mass of the string and represents a constant redshift
(thus we can analyze the orbits using the proper time and rescale by $E$ to
obtain the trajectories observed by a distant observer)
and $L$ is the conserved angular momentum per
unit mass and may be rewritten in the form $L=r^2\dot\varphi + Q\dot\varphi$.
The first term represents the free angular momentum, while the second term
represents the intrinsic angular momentum provided by the
fivebrane. This is analogous to the intrinsic angular momentum provided by
a magnetic monopole to an orbiting electron\ref\Monosca{D.G.
Boulware, L.S. Brown, R.N. Cahn, S.D. Ellis and C. Lee, Phys. Rev. D {\bf 14}
(1976) 2708.}, except that in this case the motion of the
string is restricted to its initial plane.
Note that, if $E=1$, $\dot{r}=0=\dot{\varphi}$ from \lagthree, so that the
string trivially remains stationary.
{}From now on, we assume $E>1$ for nontrivial cases.

The geodesic condition ${\cal L}=1$ can now be recast in the standard form
of General Relativity
\eqn\efpt{\dot r^2 + V^2(r)=E^2,}
where
\eqn\effpot{V^2(r)
\equiv E^2-{\lbr E^2-1\rbr\lbr r^2-a^2\rbr\over r^2 e^{4\phi}}}
is the effective potential and where $a^2\equiv {L^2\over E^2-1}-Q$.

{}From  \effpot\ it follows that for $a^2> 0$ there is a turning point at
$r_{{\rm min}}=a$ while for $a^2\leq 0$, $r_{{\rm min}}=0$.
In the physical region $r\geq r_{{\rm min}}$, $V^2$ is monotonically
decreasing so that  $\dot r$ changes sign only at the turning point.
It follows that if the test string is
directed initially away from the fivebrane (i.e. the radial component of the
initial velocity is positive), then it will spiral away to infinity.
If the string is directed towards the fivebrane, then
two qualitatively different orbits arise depending on the sign of $a^2$.
If $a^2\leq 0$, the string spirals into the fivebrane in an infinite amount
of proper time, thus never observing a singularity (the special case of $L=0$
is discussed in \refs\dkl). If $a^2 > 0$, the string spirals into the turning
point at $r_{{\rm min}}=a$, after which it swings back to infinity.

In order to explicitly solve for the two types of open orbits, we first write
down a reference-frame independent orbit equation from the expressions for
$\dot{r}$ and $\dot{\varphi}$ in \efpt\ and \com\ respectively:
\eqn\esqor{\left({\d r\over\d\varphi}\right)^2={E^2-1\over L^2}r^4
\left[1+{Q\over r^2}\right]-r^2,}
which can be simplified further using $a$ as
\eqn\essqo{\left({\d r\over\d\varphi}\right)^2={r^2\over a^2+Q}
\left(r^2-a^2\right).}
In analogy with the Keplerian analysis of Newtonian dynamics we
reparametrize by $u\equiv 1/r$ and obtain
\eqn\eskepl{\left({\d u\over\d\varphi}\right)^2={1\over a^2+Q}
\left(1-a^2 u^2\right).}
This equation of motion can be easily solved with the result
\eqna\symeom
$$\eqalignno{u&={1\over r}
={1\over a}\cos\left(\omega(\varphi-\varphi_0)\right)
\qquad\hbox{for } a^2> 0,&\symeom a\cr
u&={1\over r}
={1\over ia}\cosh\left(\omega(\varphi-\varphi_0)\right)
\qquad\hbox{for } a^2\leq 0,&\symeom b\cr}$$
where $\omega^2=\left|a^2\right|/(a^2+Q)$. The solution for the $a^2=0$ case
$u=Q^{-1/2}\left|\varphi-\varphi_\infty\right|$ arises as
a limiting case of both $\symeom{a}$ and $\symeom{b}$.

Note that in both cases the orbits are not conic! Thus we do not recover
any of the orbits of Newtonian dynamics. It is easy to see that there are
no similarities with the orbits of Schwarzschild geometry either.

A geometrical way to distinguish the above two cases from the initial
conditions
is to draw a $3$-dimensional cone in the transverse four-dimensional
space with vertex
at the string, axis along the radial direction
and half-angle ${\rm Arctan}(\sqrt{Q}/r)$. If the velocity vector (either
proper or coordinate) lies within the cone then $a^2\leq 0$. Otherwise,
$a^2> 0$.

In order to study the competition between the attractive and repulsive forces,
we compute the components of the acceleration of the test string.
The radial component is given by
\eqn\ascr{a_r={Q\over r^5 e^{6\phi}}
\left[(E^2-1)(r^2+Q)-2L^2\right]}
and the angular component is
\eqn\ascph{a_{\varphi}=\pm
{2QL \over r^5e^{6\phi}}
\left[(E^2-1)(r^2+Q)-L^2\right]^{1/2},}
where the sign is chosen according to the sign of $\dot{r}$.
Dividing by $E^2$, we can obtain the components of the
acceleration measured by a distant observer.

{}From \ascr\ we see that $a_r>0$ ($a_r<0$) if $r>r_a$ ($r<r_a$), where
$r_a^2=a^2+{L^2\over E^2-1}> r^2_{{\rm min}}$ for $L>0$.
For radial motion $r_a=0$ and $a_r>0$ always and the ``force'' is
always outward, i.e. repulsive. Usually the angular momentum signals the
existence of an outward force, but here the angular motion actually
generates an extra inward force, so that
 we have an extra inward force region
$r_a>r>r_{{\rm min}}$ for nonzero angular momentum.
The source of this attractive force is the instanton at
the center. This phenomenon is not completely unknown.
For example, in the electron-monopole case the angular motion of the
electron generates a current which interacts with the monopole charge to
generate an extra force.

It is easy to see from the acceleration that for both types of orbits we have
\eqn\forceratio{\left|{a_\varphi\over a_r}\right| > \left|{rd\varphi\over dr}
\right|,}
from which it follows that the trajectory always bends concave (i.e. inward)
with respect to the origin.
Another way to see this is to note that for the $L^2> (E^2-1)Q$ ($a^2> 0$)
case, $\Delta\varphi=\pi(1+Q/a^2)^{1/2} > \pi$, where $\Delta\varphi$ is the
total angular deviation. In fact, one can determine from the initial
conditions the number of loops the string makes around the fivebrane before
heading off to infinity. $n$ loops means
$2\pi n < \Delta\varphi \leq 2\pi ( n+1)$
which is equivalent to
\eqn\loops{\left( 1- {1\over 4n^2}\right)\left( {L^2\over E^2-1}\right)
 < Q \leq
\left(1- {1\over 4(n+1)^2}\right)\left( {L^2\over E^2-1}\right).}
Note that as $n\to\infty$, $a\to 0$ and the string spirals towards the
fivebrane. These types of ``swingshot'' orbits are
more or less analogous to electron-monopole
orbits in which the electron swings around the monopole before heading off to
infinity.

If initially the string is at $r_{{\rm min}}$ with $\dot{r}=0$, for constant
$\dot{\varphi}$ there is an unstable circular orbit. For given $E$ and
$L$ the size of this circular orbit depends only on the instanton charge $Q$
of the fivebrane source. Since $Q$ is quantized in terms of $\alpha'$,
$r_{{\rm min}}$ is discretized accordingly. This is also analogous to the
circular orbit of an electron around a monopole, in which the radius is
discretized by the quantized monopole charge.

Finally, as $r\to\infty$, $a_r\to Q(E^2-1)r^{-3}$ so that the force does not
satisfy the inverse square law but the inverse cubic law, in direct contrast
with General Relativity (a similar inverse cubic law appears in the motion of
an electron around a magnetic monopole, but only in terms of the modified
position vector\Monosca). The origin of this departure
lies partially in the presence
of the antisymmetric tensor in the string-fivebrane system, which generates
an additional repulsive force which dominates the dynamics at long distances
and precludes the existence of stable bounded orbits.

Although our analysis has been done in the transverse space of the fivebrane
only, the implications of our findings could be significant. If this structure
survives compactification, one might be able to formulate an interesting test
for string theory. Either there is no remnant of the fivebrane after
compactification if we fail to observe these new structures, or string theory
should lead to structures different from those of Newtonian dynamics at
distances longer than the Planck scale, yet at a still sufficiently short
scale (otherwise, this structure may not be observable anyhow). It is
therefore important to seriously investigate the compactifications of
the fivebrane solutions of ten-dimensional string theory.



\bigbreak\bigskip\bigskip\centerline{{\bf Acknowledgements}}\nobreak

\par\vskip.3truein
The authors would like to thank M. Duff for discussions.
This work was supported in part by NSF grant PHY89-07887 and World Laboratory
Fellowships.

\vfill\eject
\listrefs
\bye